\documentclass{moriond}

\def\Journal#1#2#3#4{{#1} {\bf #2}, #3 (#4)}

\def\NPB{{\em Nucl. Phys.} B}

\def\PRL{\em Phys. Rev. Lett.}
\def\PRD{{\em Phys. Rev.} D}

\def\EPJ{{\em Eur. Phys. J.} C}
\def\JHEP{\em JHEP}
\def\be{\begin{equation}}
\def\ee{\end{equation}}
\def\bea{\begin{eqnarray}}
\def\eea{\end{eqnarray}}

\newcommand{\Photo}{}
\newcommand{\rds}{R_{D^*}}
\newcommand{\rdp}{R_{D\pi}}
\newcommand{\pds}{p_{D^*}}
\newcommand{\dsm}{m_{D^*}}

\begin{document}
\vspace*{4cm}
\title{$R_{D^*}$ or $R_{D\pi}$: closing the theoretical gap? }
\author{J. E. Chavez-Saab, Marxil S\'anchez
and Genaro Toledo}

\address{Instituto de F\'{\i}sica, Universidad Nacional Aut\'onoma de M\'exico, AP20-364, Ciudad de M\'exico 01000, M\'exico.}

\maketitle\abstracts{
Measurements of the $R_{D^*}$ parameter remain in tension with the standard model prediction, despite recent results helping to close the gap. In this work, we revisit the standard model considerations for the prediction. We pay particular attention to the theoretical prediction considering the full 4-body decay $(B\rightarrow l\nu D^* \to l\nu D\pi)$, which introduces the longitudinal degree of freedom of the $D^*$. We show that  this does not introduce sizeable effects at the current precision. This modifies our previous finding (Phys. Rev. D 98 056014 (2018)) where a numerical bug led us to a different conclusion. Thus, the results on $R_{D\pi}$ are consistent with $\rds$, and the difference between the several values can be traced back to the form factor used and the restrictions incorporated to determine their parameters. There is still tension between the experimental world average and the most accurate theoretical estimate, leaving the possibility of presence of new physics scenarios open.}

\section{Introduction}
 One of the so-called B anomalies is observed by measuring $\rds$, which is expected to be a clean test of the lepton flavor universality, by considering the ratio between heavy and light leptons decay modes:
 \begin{equation}
 \rds \equiv\mathrm{Br}(B \to \tau\nu_\tau D^*)/ \mathrm{Br}(B \to l\nu_l D^*),
 \end{equation}
 where $ l=e, \mu$. BaBar, Belle and LHCb collaborations have conducted experiments which have consistently measured values of $\rds$ higher than the SM prediction.\\
In Table \ref{experiments}, we summarise  the experimental information and the reconstruction techniques used for the $D^*$ and the $\tau$ lepton. 
 LHCb  and early Belle results on $\rds$ relies on purely charged $D^*$ \cite{LHCb15,LHCbPRL18,LHCbPRD18,BellePRD16} (neutral B's)
while other Belle results \cite{Belle15,Belle17,Belle18} and Babar \cite{Babar12,Babar13} consider both neutral and charged ones. Recently, Belle has presented new preliminary results on $\rds$ at Moriond19 \cite{Bellemoriond19}. 
The experimental world average by the summer of 2018 quoted by the HFLAV group \cite{hflav} is $\rds=0.306 \pm 0.013\pm 0.007$. We show our own weighted average, without correlations and the uncertainties added in quadratures, for the experimental information before Moriond19 (BM19) and after Moriond19 (AM19). These results show that the world average decreased from 0.305 to 0.301.\\

In this work, we revisit the most relevant aspects of the theoretical prediction and, by taking into account that the $D^*$ is an intermediate state reconstructed from the $D\pi$ decay mode, we explore at which extend there is any  correction from the longitudinal degree of freedom by  considering $R_{D\pi}$. Our previous result \cite{chavezPRD98} is updated after finding a bug in the numerical code, bringing $R_{D\pi}$ in full agreement with $\rds$. At the end we discuss the results.
 
\begin{table}
\caption{$\rds$ as measured by several experiments. Includes preliminary results from Belle presented at Moriond 19 and our new average. (BM19: before Moriond 19, AM19: after Moriond 19)}
\label{experiments}
\begin{tabular}{| c c c|c|c| c| } 
\hline
 $\rds$  & Stat. & Syst.& $D^*$ reconstruction& $\tau$ reconstruction&Experiment \\
\hline
 0.332&0.024&0.018&$D\pi$, $D\gamma$&leptonic&BABAR12 \cite{Babar12,Babar13}\\
 0.293&0.038&0.015&$D\pi$, $D\gamma$&leptonic and hadronic&BELLE15 \cite{Belle15} \\
 0.336&0.027&0.03&$D\pi$&leptonic and hadronic&LHCb15 \cite{LHCb15}\\
 0.302&0.03&0.011&$D\pi$, $D\gamma$&leptonic and hadronic&BELLE16  \cite{BellePRD16}\\
 0.27&0.036&0.028&$D\pi$, $D\gamma$&leptonic and hadronic&BELLE17 \cite{Belle17,Belle18}\\
 0.291&0.019&0.026&$D\pi$&leptonic and hadronic&LHCb18 \cite{LHCbPRL18,LHCbPRD18}\\
\hline
0.306&0.013&0.007&&&Average(HFLAV) \cite{hflav}\\
\hline
0.305&0.012&0.009&&&Our average BM19\\
\hline
0.283&0.018&0.014&&&BELLE19(Moriond) \cite{Bellemoriond19}\\
\hline
0.301&0.011&0.008&&&Our average AM19\\
\hline
\end{tabular}
\end{table}

 \section{$\rds$}
Since the $D^*$ decay width is relatively small ($\Gamma_{D^*}$= 83.4 KeV, for the charged one), its decay process is usually considered not to play any role in the estimation of $\rds$. Thus, it is common to consider a 3-body decay $B(p_B)\rightarrow  D^*(p_{D^*},\epsilon) l(p_l) \nu (p_\nu)$, whose amplitude can be written as
\begin{equation}
M_3=\frac{G_F}{\sqrt 2}  V_{cb} 
<{D^*(\pds,\epsilon_\mu)}|J^{\lambda} | {B(p_B)}>
 l_\lambda,
\end{equation}
where $l_\lambda\equiv\bar u_l\gamma_\lambda(1-\gamma^5)v_\nu$ is the leptonic current, $V_{cb}$ is the CKM matrix element and the hadronic matrix element connecting the $B$, $D^*$ and $W$ can be parametrized in terms of four form factors, for an on-shell $D^*$. Two parametrizations of such form factors exploiting heavy quark effective theory (HQET) are the most used in the literature: \\

i) The one by Caprini {\it et al.} \cite{cln} (CLN) given by:
\begin{eqnarray}
<{D^*(\pds,\epsilon_\mu)}|J^{\lambda\mu}|{B(p_B)}>= 
\frac{2iV(q^2)}{m_B+\dsm}\epsilon^{\lambda \mu \alpha \beta}(p_B)_\alpha (p_{D^*})_\beta 
-2\dsm A_0(q^2)\frac{q^\lambda q^\mu}{q^2}\nonumber\\ 
-(m_B+\dsm)A_1(q^2)
\left( g^{\lambda \mu} - \frac{q^\lambda q^\mu}{q^2}\right) 
+\frac{A_2(q^2) q^\mu}{m_B+\dsm}
\left( (p_B+\pds)^\lambda - \frac{m_B^2-\dsm^2}{q^2} q^\lambda\right),\label{vertex}
\end{eqnarray}

\noindent where $J^\lambda =J^{\lambda\mu}\epsilon_\mu$ is the weak current, $\epsilon_\mu$ the polarization vector of the $D^*$, and $q\equiv p_B-\pds$ the transferred momentum. The form factors can be written in terms of the Universal Isgur-Wise function and parameters  obtained from a heavy quark analysis of $B^0$ decays with electron and muon products measured by the Belle collaboration~\cite{FormFactors} but not for the $\tau$. Important to notice that $A_0$ form factor is heavily suppressed for the electron and muon, while in the $\tau$ system it becomes the subdominant contribution and it is derived from $A_2$ information relying  in a HQET relation, which is valid within a 10\% accuracy. 
Using this parametrization and the experimental information a value of $\rds^{SM}=0.252\pm0.003$ is obtained \cite{fajfer12}. The error bar accounts for the uncertainties on the hadronic form factors. A similar prediction for this ratio can be obtained from an early calculation of the individual decay modes by using light front quark model (LFQM) \cite{chuan06}.\\

ii) The one by Boyd {\it et al.} \cite{bgl} (BGL) given by: 
 \begin{eqnarray}
\frac{<{D^*(\pds,\epsilon_\mu)}|J^{\mu}|{B(p_B)}>}{\sqrt{M_BM_{D^*}}}&=& 
 \epsilon^{\mu \nu \rho\sigma}\epsilon^{*\nu} \nu^{'\rho}\nu^\sigma h_V(w)
 -i(w+1)\epsilon^*_\mu h_{A_1}(w)\nonumber\\
 && i(\epsilon*\nu)\nu_\mu h_{A_2}(w)+ i(\epsilon*\nu)\nu^{'}_\mu h_{A_3}(w).
 \label{vertexBGL}
\end{eqnarray}

\noindent Using this parametrization, incorporating strong unitary restrictions for the form factors and additional experimental information on the unfolded spectrum from Belle \cite{belleffinfo}. Bigi, D. {\it et al.} \cite{bigiJHEP11} and Jaiswal, S. {\it et al.} \cite{jaiswalJHEP12} have obtained consistently higher than the ones obtained using CLN without additional theoretical restrictions. Moreover, it has been shown that by incorporating the same theoretical restrictions into CLN parametrization the value of $\rds$ increases. Berlochner, F. {\it et al.}, \cite{berlochnerPRD95} have also performed a new fit to the Belle data and imposing QCD sum rules restrictions, which is consistent with the  one obtained by Jaiswal, S. {\it et al.} \cite{jaiswalJHEP12} but with smaller uncertainty. Thus, although  all of them are consistent with each other, the most accurate value is $\rds=0.257 \pm0.003$  by Berlochner, F. {\it et al.}, \cite{berlochnerPRD95}. 
In Table \ref{theory}, we present these results for $\rds$, set in blocks to distinguish the level of the analysis and restrictions set on the form factors. Note that although labeled by CLN or BGL, they are not exactly so, as they introduce restrictions at different levels. We present a weighted average for the first block: $R^{SM}_{D^*_{AVG}}=0.258 \pm0.003$. A lattice prediction is also presented \cite{gubernari}, including LCSR, but not considered in the average as it has large uncertainties, but in agreement with it.
 
 \section{$\rdp$}
We can explore at which extent the tension of the theoretical prediction with experimental measurements is due to the fact that $D^*$ is never measured directly but through its decay into daughter particles, namely a $D\pi$ pair for the charged $D^*$  ($BR(D^*\to D\pi)=98.4\%$)  and either a $D\pi$ pair ($BR(D^*\to D\pi)=64.7\%$) or $D\gamma$  ($BR(D^*\to D\gamma)=35.3\%$) for the neutral $D^*$. Therefore, it is adequate to consider the ratio
\begin{equation}
R_{D\pi} \equiv \mathrm{Br}(B\rightarrow \tau\nu_\tau D \pi)/\mathrm{Br}(B\rightarrow l\nu_l D\pi),
\end{equation}
obtained from the full 4-body diagram shown in Fig.~\ref{4b}.
An earlier work \cite{kim} considered the full process to explore the possible effect of the $B^*$ resonance and found to be unaffected by it, obtaining a similar value for $\rdp$ as compared with $\rds$ using CLN parametrization. The purely $D^*$ longitudinal contribution ratio has also been computed \cite{fajfer12}. 
Recently, we explored the corrections that arise from the full process, corresponding to adding the longitudinal degree of freedom of the off-shell $D^*$, exhibiting the role of each contribution, transverse, longitudinal and interference, incorporating the absorptive corrections from the $D-\pi$ loops \cite{chavezPRD98}. Here, we elaborate on it and update our results after finding a numerical bug. The corrections that we focus on in this work apply only to the $D\pi$ channel and the $D\gamma$ channel is ignored throughout.
 
\begin{table}
\caption{$\rds$ and $R_{D\pi}$ as computed from several approaches. CLN: form factor parametrization using Caprini,I et al. BGL: form factors parametrization using Boyd et al. LCSR: Light cone sum rules. QCDSR: QCD sum rules }
\label{theory}
\begin{tabular}{ |c | c |c |c| } 
\hline
 $\rds$ &  Error & Approach & Reference\\
\hline
0.259&0.006&CLN& Jaiswal, S. et al. JHEP 12 (2017)060\\
0.260 & 0.008&BGL, LCSR&Bigi, D. et al. JHEP 11(2017)06\\
0.257 & 0.003&BGL, QCDSR&Berlochner, F. et al. PRD 95, 115008 (2017)\\
0.257 & 0.005&BGL&Jaiswal, S. et al. JHEP 12 (2017)060\\
\hline
0.258&0.003&& Our average\\
\hline
0.256 & 0.020&LCSR+Lattice&Gubernari, N. et al. JHEP 01(2019)150\\
0.252&&LFQM&Chuan -Hung, C. et al. JHEP 10(2006) 053\\
0.252 & 0.003&CLN&S. Fajfer, S. et al. PRD 85, 094025 (2012)\\
\hline
$R_{D\pi}$&&&\\
\hline
0.253&&Including $B^*$, $ D^*$, CLN&C. S. Kim et al. PRD 95, 013003 (2017)\\
0.253 & 0.003&$ D^*$+int., CLN&This work, corrected \\
\hline
\end{tabular}\end{table}

\subsection{Corrections from 3 to 4 body decay}\label{2}
\begin{figure}
\includegraphics[width=6cm]{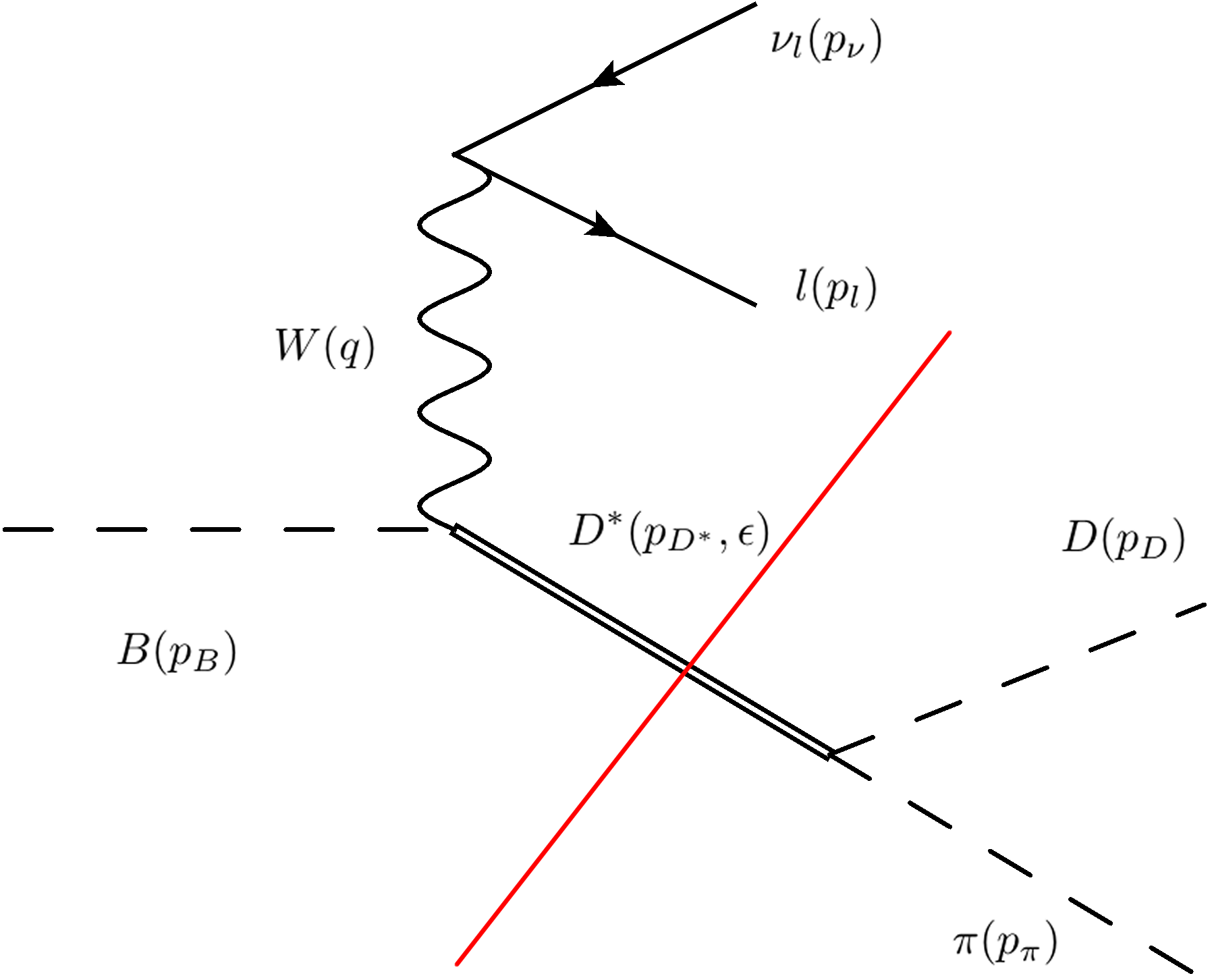}
\caption{Feynman diagram for the full $B\rightarrow l\nu_l D\pi$ decay, often thought of as independent $B\rightarrow l\nu D^*$ and $D^*\rightarrow D\pi$ decays as indicated by the red cutting line. }
\label{4b}
\end{figure}

The $B\rightarrow l\nu D^*\rightarrow l\nu D\pi$ decay can be considered as a 3-body decay with the subsequent 2-body decay processes as shown in Fig. \ref{4b}. 
The total amplitude can be written as a product of 3-body ($M_{3\mu}$) and 2-body ($M_{2\nu}$) decay amplitudes, with the polarisation tensor factored out, connected by the  $D^*$ propagator ($D^{\mu\nu}$):
$$M=M_{3\mu}D^{\mu\nu}M_{2\nu}.$$

\noindent 
The 2-body decay amplitude is $M_{2\nu}=-ig(p_D-p_\pi)_\nu$, where $p_D$ and $p_\pi$ are the momenta of the $D$ and the $\pi$, respectively, and $g$ is the $D^*-D-\pi$ coupling.

Upon considering the absorptive correction (dominated by the $D\pi$ mode), the propagator can be set in terms of the transverse and longitudinal part as follows:
\begin{equation}
\label{prop}
D^{\mu\nu}=\frac{-i T^{\mu\nu}}{\pds^2-\dsm^2+i Im\Pi_T}+\frac{i L^{\mu\nu}}{\dsm^2- i Im\Pi_L},
\end{equation}
with the corresponding projectors:
$T^{\mu\nu}\equiv g^{\mu\nu}-\frac{\pds^\mu\pds^\nu}{\pds^2}$ and $L^{\mu\nu}\equiv \frac{\pds^\mu\pds^\nu}{\pds^2},$
 where  $\pds\equiv p_D+ p_\pi$ and $\dsm$ is the mass of the $D^*$. 
 Here, the transverse correction is proportional to the full decay width, $Im\Pi_T=\sqrt{\pds^2} \Gamma_{D^*}(\pds^2)$, while the longitudinal function $Im \Pi_L = -g^2 \lambda^{1/2}(p_{D^*}^2 , m_D^2, m_\pi^2) (\frac{m_D^2 - m_\pi^2}{p_{D^*}^2})^2 / 16\pi $, which is proportional to the square mass difference  of the $D$ an $\pi$ mesons.
Since $\Gamma_{D^*}\equiv\Gamma_{D^*}(\dsm^2)$ is relatively small, the relevant contribution to the transversal term is just around $\pds^2=\dsm^2$, and a narrow width approximation can be used. This allows us to rewrite the transversal part of the squared amplitude as
\begin{equation}\label{Tamp}
|M_T|^2=M_{3\mu}M_{3\alpha}^*T^{\mu\nu}T^{\alpha\beta}M_{2\nu}M_{2\beta}^*\frac{\pi \delta(\pds^2-\dsm^2)}{\dsm\Gamma_{D^*}}.
\end{equation}
The delta function forces the transverse part of the $D^*$ to remain on-shell. Therefore, the transverse part of the $D^*$ propagator is equivalent to the 3-body decay.
On the other hand, the longitudinal part of the propagator gives place to two new terms in the squared amplitude (one purely longitudinal and one of interference), modulated by a dimensionless mass-difference parameter $\Delta^2\equiv(m_D^2-m_\pi^2)/\dsm^2=0.86$, that cannot be accounted for in the $B\rightarrow l\nu_lD^*$ process: 
 The pure longitudinal part of the squared amplitude can be written as
\begin{equation}
|M_L|^2=|M_{3\mu}L_{\mu\nu}M_2^\nu|^2 \frac{1}{\dsm^4+(Im\Pi_L)^2},
\end{equation}
while the interference is proportional to
\begin{equation}
\frac{(\pds^2-\dsm^2+ i \dsm\Gamma_{D^*})}{\dsm^2-i Im\Pi_L}T^{\mu_1\nu_1}L_{\mu_2\nu_2}
 \frac{\pi}{ \dsm\Gamma_{D^*}} \delta(\pds^2-\dsm^2).
\end{equation}

An earlier estimate \cite{kim} differs from ours (in the limit of $Im\Pi_L=0$) by the term proportional to $i\dsm \Gamma_{D^*}$ traced back to the form of their longitudinal part. The interference upon integration is forced by the transversal part to be on-shell, where this term is the only not null contribution. Notice that this imaginary term makes a real contribution as both the leptonic tensor and the $B-D^*-W$ vertex carry also an imaginary term.

In calculating the new decay widths, both the transverse and interference parts of the squared amplitudes have been integrated as being on-shell through the narrow width approximation, while the longitudinal part is integrated in the 4-body phase space within a window  $\pds^2=(\dsm \pm \delta)^2$. We have explored values for $\delta$ around $\Gamma_{D^*}/2$ to 1 MeV and found the final result for $R_{D\pi}$ to be unaffected by the particular choice.\\
In Table \ref{table1}, we show our updated $R_{D\pi}$ for the electron and the muon as each part is added, namely, transversal, longitudinal and interference parts. Notice that, due to the cancellation of global factors in the ratio, $R_{D\pi}$ has a much higher precision than the individual branching ratios. We note that the pure longitudinal contribution to the branching ratio is the same for the light leptons. On the other hand, the relative size of the interferences turns out to be negligible at the current precision, however we have added more significant figures to exhibit their role. The uncertainty on $R_{D\pi}$ is similar to $\rds$ and comes from the uncertainties on the measurement of the form factors in the CLN parametrization for which we have used results published by Belle \cite{FormFactors} without further restrictions.

\begin{table}
\caption{Contribution to the branching ratio of the transversal, longitudinal and interference parts of the amplitude for all three lepton flavors. Quantities are given in percentage. The last two rows shows the value of $R_{D\pi}$ as each contribution is added subsequently from left to right for $e$ and $\mu$.}
\label{table1}
\begin{tabular}{ |c|c c c |} 
 \hline
& Transversal  & Longitudinal & Interference\\
\hline
Electron & $4.6(3)$  & $5.0(3)\times10^{-6}$ & $3.2(2)\times 10^{-3}$\\
Muon & $4.6(3)$ & $5.0(3)\times 10^{-6}$ & $3.2(2)\times10^{-3}$ \\
Tau & $1.16(8)$ & $1.1(6)\times10^{-6}$ &$ 6.4(2)\times10^{-4}$ \\
\hline
\hline
$R_{D\pi}^e$ &0.25221&0.25221&0.25217\\
$R_{D\pi}^\mu$ &0.25330&0.25330&0.25327\\
\hline
\end{tabular}
\end{table}

\section{Discussion}
The accurate theoretical and experimental information on $\rds$ are importan to elucidate if indeed there is evidence of the violation of lepton flavour universality. In this work, after a short review of the experimental and theoretical status, we have elaborated on the longitudinal correction from the $D^*$ propagator to $R_{D\pi}$. We have incorporated the absorptive corrections and found that the interference with the transversal part is the most significant contribution but makes no difference as compared to $\rds$ at the current precision. Our previous result \cite{chavezPRD98} is updated after finding a bug in the numerical code, bringing $R_{D\pi}$ in full agreement with $\rds$ and the difference between the several values can be traced back to the form factor used and the restrictions incorporated to determine their parameters. The experimental world average and the most accurate theoretical estimate are still in tension leaving the possibility of presence of new physics scenarios open \cite{straub}.

A full 4-body description would require to consider the most general structure for the four body decay. In our description, terms proportional to $\pds^\mu$ coupled to the longitudinal part of the $D^*$ propagator are absent. Additional longitudinal terms in the $D^*-D-\pi$ vertex have not been considered either. Since the form factors have been derived from Belle data \cite{FormFactors} without including these terms, a new analysis should be necessary.\\
Contributions from the scalar resonances may be part of the internal process, which were taken into account in the background analysis by the experiments. A lack of information on these states make theoretical descriptions to be rough estimates \cite{marxil}. Also radiative corrections in the pseudo-scalar and neutral vector mesons systems are expected to be important at the few percent level in the branching ratios \cite{radcor,radcor2,radcor3}. Thus, upcoming improvements on the experimental side will require to account for them in the theoretical prediction.\\

Acknowledgments.
G. T. Thanks the organizers for a wonderful conference and partial support. This work was supported in part by a PIIF-IFUNAM project and CONACyT project 252167ÐF.

\end{document}